\documentclass[aps, pra, twocolumn, showpacs]{revtex4-1}%

\pdfoutput=1

\usepackage[latin1]{inputenc}
\usepackage[english]{babel}
\usepackage[T1]{fontenc}
\usepackage[dvips]{graphicx}
\usepackage{amsmath}
\usepackage{amsfonts}
\usepackage{amssymb}
\usepackage{epsfig}
\usepackage{graphicx,color}
\usepackage{dcolumn}

\newcommand{\tabref}[1]{Table~\ref{#1}}

\newcommand{\eqnref}[1]{Eq.~(\ref{#1})}

\begin{document}

\title{Scaling Invariance of Density Functionals}
\author{L\'{a}zaro~Calder\'{i}n}
\email{calderin@psu.edu}
\affiliation{Materials Research Institute and Research Computing and Cyberinfrastructure, 
             The Pennsylvania State University,
             University Park, PA, USA}
\date{\today}
\begin{abstract}
 Based on the homogeneity ($F[n_{\lambda m}]=\lambda^{p(m)}F[n]$) and invariance ($F[n_{\lambda m_0}]=F[n]$) properties of a functional of the electron density under uniform scaling of the coordinates in the density ($n_{\lambda m}(\mathbf{r})=\lambda^{m} n(\lambda\mathbf{r}),\,\lambda\in\mathbb{R}^+,\, m\in\mathbb{R})$, it is proven that homogeneity implies invariace and therefore all homogeneous scaling functionals have the representation $F[n]=\frac{m-m_0}{p(m)} \int_V\,\frac{\delta F[n]}{\delta n(\mathbf{r})}\,n(\mathbf{r})\,d^3r$. 
Also, the homogeneity ($p(m)$) and invariant ($m_0$) degrees of density functionals related to the Kohn-Sham theory are calculated. Besides, it is shown that the functional density and the electron density itself satisfy the general equation representing the local scaling invariance of a functional: $\lambda \frac{d}{d\lambda} f([n_{\lambda m_0}],\mathbf{r},\mathbf{r'}) 
= 
  \sum_{i=1}^3 \frac{d}{d x_i}   \left[ x_i  f([n_{\lambda m_0}],\mathbf{r},\mathbf{r'}) \right]
+ \sum_{j=1}^3 \frac{d}{d x_j'} \left[ x_j' f([n_{\lambda m_0}],\mathbf{r},\mathbf{r'}) \right]
$. 
The equation simplifies for cases where the functional density depends only on the density and/or its gradient, and general forms of the solutions are provided. In particular the non-interacting kinetic energy density is shown to take the form 
$t_s(n,\nabla n)= n(\mathbf{r})^{3} g\left[
\frac{\partial_{x_1} n(\mathbf{r})}{n(\mathbf{r})^2},
\frac{\partial_{x_2} n(\mathbf{r})}{n(\mathbf{r})^2},
\frac{\partial_{x_3} n(\mathbf{r})}{n(\mathbf{r})^2}\right]$
under such conditions.
\end{abstract}
\maketitle

\section{Introduction}
Density functional theory (DFT) \cite{hk64} in its Kohn-Sham (KS) \cite{PhysRev.140.A1133} form continues to dominate the landscape of ab-initio simulations of atoms, molecules, solids and liquids; expanding applications in condensed-matter physics, chemistry, biology, materials science and device engineering.

However, the sucess of KSDFT comes also with serious drawnbacks due to the necessary use of approximated exchange-correlation functional ($E_{xc}$) \cite{PhysRev.140.A1133}. Similarly, applications of the so-called orbital free methods \cite{OF} are also seriously limited by the approximations introduced to the KS kinetic-energy functional $T_s$. A way to improve the approximations for both $E_{xc}$ and $T_s$ is by investigating their scaling properties.

Although the scaling properties of density functionals have been investigated by several groups \cite{Sham70,PhysRevA.32.2010, DreizlerGross1990Cp4.7,LiuParrCPL278,PhysRevLett.97.223002,PhysRevA.77.022504,PhysRevA.86.032510}, important questions still remain to be answered. Namely, the global and local invariance of density functionals under homogeneous scaling of the coordinates in the electron density $n(\mathbf{r})$, which are the subjects of this study.

We shall start by introducing the concept of global invariance of a density functional, describe how to find the scaling that leaves invariant a given functional and illustrate it with an example. Analogously, we will define local invariance of a density functional, investigate its consequences and find general formulas for special cases. 

But before all that, we need to remember that in the context of KSDFT the total energy density functional $E_t[n]$ is defined as
\begin{align}
 E_t[n] = T_s[n] + E_H[n] + E_{xc}[n] + E_{ext}[n],
\end{align}
where $T_s[n]$ is the kinetic energy density functional of the non-interacting system;
$E_H[n]$ is the Hartree density functional
\begin{align}
 E_H[n]=\frac{1}{2} \int_V \int_V \frac{n(\mathbf{r}) n(\mathbf{r'})}{|\mathbf{r}-\mathbf{r'}|} d^3r d^3r';
\end{align}
$E_{xc}[n]$ is the exchange-correlation energy density functional;
and $E_{ext}[n]$ is the external energy density functional for a given external potential $v_{ext}(\mathbf{r})$, that is 
\begin{align}
 E_{ext}[n] = \int_V v_{ext}(\mathbf{r})\, n(\mathbf{r})\, d^3r.
\end{align}

\section{Global Scaling Invariance}
A density functionals can be written in general as 
\begin{equation}
 F[n]=\int_V\int_V \, f([n],\mathbf{r},\mathbf{r'}) \, d^3r d^3r',
\end{equation}
where $f([n],\mathbf{r},\mathbf{r'})$ is an energy or functional density that goes to zero at the boundary of the volume $V$, which is taken as the whole space.
We are interested on the general homogenous scaling of the coordinates in the electron density
\begin{equation}
\label{eq:nscaling}
 n_{\lambda m}(\mathbf{r})=\lambda^m n(\lambda\mathbf{r}),\,\lambda\in\mathbb{R}^+,\, m\in\mathbb{R}.
\end{equation}
A density functional $F[n]$ is homogeneous of degree $p(m)$ if under such density scaling the functional scales as \cite{citeulike:6901027}
\begin{equation}
 F[n_{\lambda m}]=\lambda^{p(m)} F[n];
\label{eq:Fscaling}
\end{equation}
and, it is {\it globally invariant} if for some value of the scaling exponent or degree $m_0$ we have $p(m_0)=0$, and therefore
\begin{equation}
 F[n_{\lambda m_0}] = F[n].
\label{eq:Finvariance}
\end{equation}

A consequence of the homogeneity property is that the functional can be written in terms of its functional derivative as
\begin{equation}
 F[n]=\frac{1}{p(m)} \int_V\, \frac{\delta F[n]}{\delta n(\mathbf{r})}
\left( {m\, n(\mathbf{r} ) + \mathbf{r} \cdot \nabla n(\mathbf{r})} \right) d^3r;
\end{equation}
while the global invariance condition yields
\begin{equation}
\int_V\, \frac{\delta F[n]}{\delta n(\mathbf{r})} \left( {m_0\, n(\mathbf{r} ) + \mathbf{r} \cdot \nabla n(\mathbf{r})} \right) d^3r=0;
\end{equation}
as shown by parametric derivation respect to $\lambda$ of \eqnref{eq:Fscaling} and \eqnref{eq:Finvariance}, respectively.
Therefore, any $p(m)$-homogeneous and $m_0$-invariant functional can be written just in terms of its functional derivative and the density as
\begin{equation}
 F[n]=\frac{m-m_0}{p(m)} \int_V\,\frac{\delta F[n]}{\delta n(\mathbf{r})}\,n(\mathbf{r})\,d^3r.
\label{eq:intrep}
\end{equation}

Notice that $p(m)$ is in fact the linear function $p(m)=q*m+k$ with $q$ and $k$ also real numbers. Hence, if $q$ is not zero, there is always a value $m_0=-k/q$ for which the a homogeneous scaling functional is left invariant. 
Therefore, every homogeneous scaling invariant functional under homogeneous scaling of the coordinates in the electron density, is also invariant of some degree.
Consequently, all homogeneously scaling functionals have the form of \eqnref{eq:intrep}.


\subsection{Global scaling invariance of KS density functionals}
The scaling properties of $T_s[n]$ under general homogeneous scaling were investigated in details in \cite{PhysRevA.86.032510}, where it was also proven that $T_s[n]$ is global invariant under the scaling \eqnref{eq:nscaling} for $m_0=1$.

For the Hartree potential we have
\begin{align}
 E_H[n_{\lambda}]=&
\frac{1}{2} \int_V \int_V \frac{n_{\lambda}(\mathbf{r})\, n_{\lambda}(\mathbf{r'})}{|\mathbf{r}-\mathbf{r'}|}d^3r d^3r'
\\\nonumber
=&
\frac{\lambda^{2m-5}}{2} \int_V \int_V \frac{n(\lambda\mathbf{r}) n(\lambda\mathbf{r'})}{|\lambda\mathbf{r}-\lambda\mathbf{r'}|}
\lambda^{6} d^3r d^3r';
\end{align}
given that the integration volume $V$ is all the space. Therefore $p(m)=5m-2$ and for $m_0=5/2$ we get
\begin{equation}
E_H[n_{\lambda m_0}]=E_H[n].
\end{equation}

Similarly, it can be also proven that $E_{ext}[n]$ for an atom at the origin and the total number of electrons $N_e[n]=\int_V\,n(\mathbf{r})d^3r$ are invariant under the scaling \eqnref{eq:nscaling} for $m_0=2$ and $m_0=3$, respectively. 

\tabref{tab:mindex} lists, for all those KSDFT-related homogeneous functionals, their homogeneous scaling and invariance degrees. All of those functionals admit the integral representation \eqnref{eq:intrep}, including $T_s[n]$. In contrast, $E_{xc}[n]$ \cite{Levy85} and, generally $E_{ext}[n]$, are not invariant under homogeneous coordinate scaling of the density and therefore have no invariance.
%
%
\begin{table}
\caption{\label{tab:mindex}Homogeneous ($p(m)$) and invariance ($m_0$) degrees  of related KS density functionals.}
\begin{ruledtabular}
\begin{tabular}{ccccc}
$F[n]$    & $T_s[n]$     & $E_H[n]$ & $N_e[n]$ & $E_{ext}[n]$   \\\hline
$p(m)$ & $m-1$ & $2m-5$ & $m-3$ & $m-2$ \\
$m_0$ & $1$ & $5/2$ & $3$ & $2$ \\
\end{tabular}
\end{ruledtabular}
\end{table}

\section{Local Scaling Invariance}
%
If $F[n]$ is global invariant under a homogeneous scaling of the density, then the functional density scales as 
\begin{equation}
 f([n_{\lambda m_0}],\mathbf{r},\mathbf{r'}) = \lambda^6 f([n],\lambda\mathbf{r},\lambda\mathbf{r'})  
\end{equation}
or if $f$ depends only once on the coordinates then
\begin{equation}
 f([n_{\lambda m_0}],\mathbf{r}) = \lambda^3 f([n],\lambda\mathbf{r});  
\end{equation}
which means that for arbitrary integration limits it is true that
\begin{widetext}
\begin{equation}
\int_{a_1}^{b_1} \int_{a_2}^{b_2} \int_{a_3}^{b_3}
\int_{a_1'}^{b_1'} \int_{a_2'}^{b_2'} \int_{a_3'}^{b_3'} f([n],\mathbf{r},\mathbf{r'})\, d^3r\, d^3r'
=
\int_{a_1/\lambda}^{b_1/\lambda} \int_{a_2/\lambda}^{b_2/\lambda} \int_{a_3/\lambda}^{b_3/\lambda} \int_{a_1'/\lambda}^{b_1'/\lambda} \int_{a_2'/\lambda}^{b_2'/\lambda} \int_{a_3'/\lambda}^{b_3'/\lambda}
 f([n_{\lambda m_0}],\mathbf{r},\mathbf{r'})\,  d^3r\, d^3r'.
\label{eq:locinvcond}
\end{equation}
\end{widetext}
 
In other words, the integral for finite and arbitrary integration limits must be invariant under the scaling, taking also  into account the corresponding changes in the integration limits. This os the {\it local scaling invariance} condition, which as we will see, although simple, has far reaching consequences. 
 
It follows that the parametric derivative of the right-hand side of the local scaling invariance condition \eqnref{eq:locinvcond}, respect to the scaling parameter $\lambda$, must be zero for all possible values of $\lambda$.
Taking such a derivative using Leibniz's rule for the parametric derivative of an integral,\cite{Leibnizrule} and taking into account that the integration limits are arbitrary, we get the following general equation involving the functional density $f([n],\mathbf{r},\mathbf{r'})$ and the density $n(\mathbf{r})$ 
\begin{multline}
\lambda \frac{d}{d\lambda} f([n_{\lambda m_0}],\mathbf{r},\mathbf{r'}) 
= 
  \sum_{i=1}^3 \frac{d}{d x_i}   \left[ x_i  f([n_{\lambda m_0}],\mathbf{r},\mathbf{r'}) \right]
\\+ \sum_{j=1}^3 \frac{d}{d x_j'} \left[ x_j' f([n_{\lambda m_0}],\mathbf{r},\mathbf{r'}) \right].
\label{eq:local-invariance-1}
\end{multline}
This is a fundamental equation which the functional density of any functional of invariant degree $m_0$ must obey.

It is impossible to proceed without further assumptions on the depencency of the functional density on the electron density, its derivatives and coordinates. One possibility is that the functional density
$f$ is explicitly dependendent only on the density and its gradient $\nabla n$, that is
\begin{equation}
 f([n_{\lambda m}],\mathbf{r},\mathbf{r'})=f(n_{\lambda m},\nabla n_{\lambda m}, \mathbf{r},\mathbf{r'}),
\end{equation}
in that case we can write \eqnref{eq:local-invariance-1} as
\begin{widetext}
\begin{multline}
\lambda  
\left[
\frac{\partial f(n_{\lambda m_0},\partial n_{\lambda m_0},\mathbf{r},\mathbf{r'})}{\partial n_{\lambda m_0}(\mathbf{r})}  
\frac{d n_{\lambda m_0}(\mathbf{r})}{d\lambda}
+
\sum_{i=1}^3
\frac{\partial f(n_{\lambda m_0},\partial n_{\lambda m_0},\mathbf{r},\mathbf{r'})}{\partial (\partial_i n_{\lambda m_0}(\mathbf{r}))}  
\frac{d \partial_i n_{\lambda m_0}(\mathbf{r})}{d\lambda}
\right]
\\
+
\lambda  
\left[
\frac{\partial f(n_{\lambda m_0},\partial n_{\lambda m_0},\mathbf{r},\mathbf{r'})}{\partial n_{\lambda m_0}(\mathbf{r'})}  
\frac{d n_{\lambda m_0}(\mathbf{r'})}{d\lambda}
+
\sum_{i=1}^3
\frac{\partial f(n_{\lambda m_0},\partial n_{\lambda m_0},\mathbf{r},\mathbf{r'})}{\partial (\partial_i n_{\lambda m_0}(\mathbf{r'}))}  
\frac{d \partial_i n_{\lambda m_0}(\mathbf{r'})}{d\lambda}
\right]
\\=
 \sum_{i=1}^3 \frac{\partial f(n_{\lambda m_0},\partial n_{\lambda m_0},\mathbf{r},\mathbf{r'})}{\partial n_{\lambda m_0}(\mathbf{r})}
x_i\partial_i n_{\lambda m_0}(\mathbf{r})  
+
  \sum_{i=1}^3 \sum_{j=1}^3 
\frac{\partial f(n_{\lambda m_0},\partial n_{\lambda m_0},\mathbf{r},\mathbf{r'})}{\partial (\partial_i n_{\lambda m_0}(\mathbf{r}))}
x_j\partial_j \partial_i n_{\lambda m_0}(\mathbf{r})  
\\
+
  \sum_{i=1}^3 \frac{\partial f(n_{\lambda m_0},\partial n_{\lambda m_0},\mathbf{r},\mathbf{r'})}{\partial n_{\lambda m_0}(\mathbf{r'})}
x_i\partial_i n_{\lambda m_0}(\mathbf{r'})  
+
  \sum_{i=1}^3 \sum_{j=1}^3 
\frac{\partial f(n_{\lambda m_0},\partial n_{\lambda m_0},\mathbf{r},\mathbf{r'})}{\partial (\partial_i n_{\lambda m_0}(\mathbf{r'}))}
x'_j\partial_j \partial_i n_{\lambda m_0}(\mathbf{r'})  
\\
+
\nabla_{\mathbf{r}}\cdot \left[\mathbf{r} f(n_{\lambda m_0},\partial n_{\lambda m_0},\mathbf{r},\mathbf{r'}) \right]
+
\nabla_{\mathbf{r'}}\cdot \left[\mathbf{r'} f(n_{\lambda m_0},\partial n_{\lambda m_0},\mathbf{r},\mathbf{r'}) \right];
\label{eq:local-invariance-3}
\end{multline}
which, by mean of the identities
\begin{equation}
 \lambda \frac{d n_{\lambda m}(\mathbf{r})}{d\lambda} = m\, n_{\lambda m}(\mathbf{r}) + \sum_{i=1}^3  x_i \partial_i n_{\lambda m}(\mathbf{r})
\;\;\text{and}\;\;
\lambda \frac{d \partial_i n_{\lambda m}(\mathbf{r})}{d\lambda} 
=
(m+1)\, \partial_i n_{\lambda m}(\mathbf{r}) + \sum_{j=1}^3 x_j \partial_j \partial_i n_{\lambda m}(\mathbf{r}), 
\end{equation}
is reduced to
\begin{multline}
m_0
\left[
\frac{\partial f(n_{\lambda m_0},\partial n_{\lambda m_0},\mathbf{r},\mathbf{r'})}{\partial n_{\lambda m_0}(\mathbf{r})}  
n_{\lambda m_0}(\mathbf{r})
+
\frac{\partial f(n_{\lambda m_0},\partial n_{\lambda m_0},\mathbf{r},\mathbf{r'})}{\partial n_{\lambda m_0}(\mathbf{r'})}  
n_{\lambda m_0}(\mathbf{r'})
\right]\\
+
(m_0+1)
\sum_{i=1}^3
\left[
\frac{\partial f(n_{\lambda m_0},\partial n_{\lambda m_0},\mathbf{r},\mathbf{r'})}{\partial (\partial_i n_{\lambda m_0}(\mathbf{r}))}  
\partial_i n_{\lambda m_0}(\mathbf{r})
+
\frac{\partial f(n_{\lambda m_0},\partial n_{\lambda m_0},\mathbf{r},\mathbf{r'})}{\partial (\partial_i n_{\lambda m_0}(\mathbf{r'}))}  
\partial_i n_{\lambda m_0}(\mathbf{r'})
\right]
\\
=
\nabla_{\mathbf{r}}\cdot \left[\mathbf{r} f(n_{\lambda m_0},\partial n_{\lambda m_0},\mathbf{r},\mathbf{r'}) \right]
+
\nabla_{\mathbf{r'}}\cdot \left[\mathbf{r'} f(n_{\lambda m_0},\partial n_{\lambda m_0},\mathbf{r},\mathbf{r'}) \right],
\label{eq:fexplicitandnonlocal}
\end{multline}
\end{widetext}
where $m_0$ is the invariant degree or exponent of $F[n]$, and the gradient operator $\nabla$ only acts on the explicit dependence of $f$ on the coordinates.

A quick check of this formula can be done for those explicit functionals of the density in \tabref{tab:mindex}, and the known espressions of von Weizs\"acker \cite{vW35} and Thomas-Fermi (TF) \cite{thomas,fermi} for $T_s[n]$; which have invariant exponents of $1$ and $9/5$ respectively. Although such invariant degree for TF is just apparent \footnote{In fact the Thomas-Fermi functional scales as $T_s$ and has also an invariant degree of one (see L. Calder\'in, arXiv:1404.2970) \label{note1}}.

For functional densities that do not depend explicitly on the coordinates the right hand side of \eqnref{eq:fexplicitandnonlocal} is further simplified to $3\,f$. Besides, if the functional density depends only on the density then we have the equation
\begin{equation}
m_0
\frac{\partial f(n_{\lambda m_0}(\mathbf{r}))}{\partial n_{\lambda m_0}(\mathbf{r})}  
n_{\lambda m_0}(\mathbf{r})
=
3 f(n_{\lambda m_0}(\mathbf{r})),
\label{eq:fexplicitandnonlocal-density}
\end{equation}
with solution
\begin{equation}
 f(n)=C\,n(\mathbf{r})^{3/m_0},\qquad C=constant;
\label{eq:sln1}
\end{equation}
while if the functional density depends only on the gradient of the density then the equation turns out to be
\begin{multline}
(m_0+1)
\sum_{i=1}^3
\frac{\partial f(\partial n_{\lambda m_0}(\mathbf{r}))}{\partial (\partial_i n_{\lambda m_0}(\mathbf{r}))}  
\partial_i n_{\lambda m_0}(\mathbf{r})
\\
=
3 f(\partial n_{\lambda m_0}(\mathbf{r})),
\label{eq:fexplicitandnonlocal-grad}
\end{multline}
with general solutions of the form
\begin{equation}
f(\nabla n)= (\partial_{x_1} n(\mathbf{r}))^{\frac{3}{m_0+1}} g\left[
\frac{\partial_{x_2} n(\mathbf{r})}{\partial_{x_1} n(\mathbf{r})},
\frac{\partial_{x_3} n(\mathbf{r})}{\partial_{x_1} n(\mathbf{r})}\right],
\label{eq:sln3}
\end{equation}
where $g$ is a function only of the ratios of partial derivatives.

Examples of functional densities that only depend on the electron density andt must obey \eqnref{eq:sln1} are: the number of electron density $n(\mathbf{r})$ (in $N_e[n]=\int_V n(\mathbf{r}) d^3r$) with $m_0=3$; and the Thomas-Fermi kinetic energy density which is proportional to $n(\mathbf{r})^{5/3}$ with $m_0=9/5$ (see \footnotemark[\value{footnote}]). In contrast, there are not functionals that depend only on the gradient of the electron density in DFT.

Another common situation is that the functional or energy densities depend only on the density and its gradient.
The equation to solve for such cases is
\begin{multline}
m_0
\frac{\partial f(n_{\lambda m_0}(\mathbf{r}))}{\partial n_{\lambda m_0}(\mathbf{r})}  
n_{\lambda m_0}(\mathbf{r})\\
+
(m_0+1)
\sum_{i=1}^3
\frac{\partial f(\partial n_{\lambda m_0}(\mathbf{r}))}{\partial (\partial_i n_{\lambda m_0}(\mathbf{r}))}  
\partial_i n_{\lambda m_0}(\mathbf{r})
\\
=
3 f(n_{\lambda m_0}(\mathbf{r}));
\label{eq:fexplicitandnonlocal-density-grad}
\end{multline}
with general solutions of the form
\begin{equation}
f(n,\nabla n)= n(\mathbf{r})^{\frac{3}{m_0}} g\left[
\frac{\partial_{x_1} n(\mathbf{r})}{n(\mathbf{r})^{\frac{m_0+1}{m_0}}},
\frac{\partial_{x_2} n(\mathbf{r})}{n(\mathbf{r})^{\frac{m_0+1}{m_0}}},
\frac{\partial_{x_3} n(\mathbf{r})}{n(\mathbf{r})^{\frac{m_0+1}{m_0}}}\right],
\label{eq:sln2}
\end{equation}
where $g$ is in principle an arbitrary function, but only the ratios of the gradient component of the density to a given power of the density. 

Therefore, according to \eqnref{eq:sln2} the non-interacting kinetic energy density $t_s$, which again has an invariant exponent of one, has the general form
\begin{equation}
t_s(n,\nabla n)= n(\mathbf{r})^{3} g\left[
\frac{\partial_{x_1} n(\mathbf{r})}{n(\mathbf{r})^2},
\frac{\partial_{x_2} n(\mathbf{r})}{n(\mathbf{r})^2},
\frac{\partial_{x_3} n(\mathbf{r})}{n(\mathbf{r})^2}\right],
\label{eq:ts}
\end{equation}
if $t_s$ only depends on the density and its gradient. 

An example of a non-interacting kinetic energy density that depends explicitly only on the electron density and its partial derivatives respect to the coordinates is the von Weizs\"acker energy density, which is proportional to $(\nabla n(\mathbf{r}))^2/n(\mathbf{r})$ and therefore follows the general form given by \eqnref{eq:ts}. 

Moreover, it is important to notice that when $f$ also depends on the coordinates once there is a general solution of \eqnref{eq:fexplicitandnonlocal} of the form
\begin{widetext}
\begin{equation}
f(n,\nabla n,\mathbf{r})= 
n(\mathbf{r})^{\frac{3}{m_0}} g_1\left[
\frac{\partial_{x_1} n(\mathbf{r})}{n(\mathbf{r})^{\frac{m_0+1}{m_0}}},
\frac{\partial_{x_2} n(\mathbf{r})}{n(\mathbf{r})^{\frac{m_0+1}{m_0}}},
\frac{\partial_{x_3} n(\mathbf{r})}{n(\mathbf{r})^{\frac{m_0+1}{m_0}}},
{x_1} n(\mathbf{r})^{1/m_0},
{x_2} n(\mathbf{r})^{1/m_0},
{x_3} n(\mathbf{r})^{1/m_0}
\right];
\label{eq:slng}
\end{equation}
\end{widetext}
and a corresponding expression for double dependency of the functional density on the coordinates is also possible to be obtained.

\section{Closing remarks}
%
Practical applications of Density Functional Theory are possible through approximations to the non-interacting kinetic energy functional and/or the exchange-correlation functional. Such approximations are mainly based on scaling properties of the functionals. Here have investigated the global and local invariance of functionals under homogeneous scaling of the electron density (\eqnref{eq:nscaling}). 

We showed that homogeneous functionals are also invariants of some degree and can be written in a simple form  (\eqnref{eq:intrep}), that is as an integral of the first functional derivative respect to the electron density times the electron density. We also worked out the invariance degrees of the KSDFT related functionals (\tabref{tab:mindex}).

Furthermore, we found that the functional densities of invariant functionals satisfy a local equation (\eqnref{eq:local-invariance-1}), which yields their general analytical forms. Such equation is simplified for the case of dependency of the energy functional only on the density as well as the density and its gradient (\eqnref{eq:fexplicitandnonlocal}). We provided the general forms of the solutions for those special cases (\eqnref{eq:sln1}, \eqnref{eq:sln2} and \eqnref{eq:sln3}). Finally, the general form of non-interacting kinetic energy density is found (\eqnref{eq:ts}) assuming dependency only on the density and its gradient.



%
\end{document}